\documentclass[aps,prc,reprint]{revtex4-1}
\UseRawInputEncoding
\usepackage{amssymb,amsbsy,amsfonts,amsmath}
\usepackage{epsfig,color}
\usepackage{epstopdf}
\usepackage[citecolor=blue,colorlinks=true,linkcolor=blue,urlcolor=blue]{hyperref}
\usepackage{bm}
\usepackage{braket}
\usepackage{amsmath}

\begin{document}
\title{Multiphoton fusion of light nuclei in intense laser fields}
\author{Binbing Wu$^{1}$}
\author{Zhengfeng Fan$^{1,3}$}
\email{fan\underline{_}zhengfeng@iapcm.ac.cn}
\author{Difa Ye$^{1}$}
\author{Tao Ye$^{1}$}
\author{Congzhang Gao$^{1}$}
\author{Chengxin Yu$^{1}$}
\author{Xuefeng Xu$^{1}$}
\author{Cunbo Zhang$^{1}$}
\author{Jie Liu$^{2,3}$}
\email{jliu@gscaep.ac.cn}
\affiliation{$^{1}$Institute of Applied Physics and Computational Mathematics, Beijing 100088,People's Republic of China
\\$^{2}$Graduate School, China Academy of Engineering Physics, Beijing 100193, People's Republic of China
\\$^{3}$CAPT, HEDPS, and IFSA Collaborative Innovation Center of MoE, Peking University, Beijing 100871, People¡¯s Republic of China}
\date{\today}
\begin{abstract}
We investigate the fusion cross sections of light nuclei in the presence of linearly polarized intense laser fields. By combining the Coulomb-Volkov solutions with the complex spherical square-well nuclear potential, we derive an explicit formulation of the multiphoton cross section in a self-consistent manner. Our analysis is specifically focused on deuteron-triton (DT) and proton-boron (p~$^{11}{\rm B}$) fusion reactions, both of which have garnered widespread attention. The theoretical results reveal that, under conditions of longer laser wavelengths and lower incident
particle kinetic energies, a few thousands of photons can participate in the fusion reactions, resulting in a substantial enhancement of fusion cross sections by almost ten orders of magnitude. We elucidate the multiphoton mechanism underlying
these findings and discuss their implications.
\end{abstract}
\maketitle

\section{Introduction}

The investigation of light-matter interactions holds a prominent position in physics research \cite{Joachain2012,Liu:2014}. The identification of multiphoton ionization \cite{Damon1963,Manus1991} in
atoms and molecules exposed to intense laser fields has unveiled the nonperturbative regime in strong laser fields \cite{Joachain2012}. This discovery has paved the way for diverse nonperturbative phenomena, including higher-order above-threshold ionization \cite{Paulus1994}, nonsequential double ionization \cite{Walker1994}, and high-order harmonic generation \cite{McPherson1987,Ferray1988}.

In recent years, advancements in the chirped-pulse amplification (CPA) technique have increased laser intensity
to $10^{23}{\rm W/cm^2}$ \cite{Yoon}. Furthermore, the Extreme Light Infrastructure for Nuclear Physics (ELI-NP) in
Europe is poised to generate high-intensity lasers at $10^{25}{\rm W/cm^2}$ in the near future \cite{http:eli}.
These strong laser fields impact not only atomic and molecular processes \cite{Joachain2012,Liu:2014} but also
nuclear processes \cite{Fu2022,Wang:2021,QiPRL2023,Delion:2017,Qi:2019,Ghinescu:2020,Palffy:2020,Qi:2020,Queisser:2019,Lv:2019,Wang:2020,Liu:2021,Kohlfu:2021,Lv:2022,
Qi:2022,Wu20221,Bekx20221,Bekx20222,liu2022}. For instance, recent theoretical studies reveal that intense laser fields can excite $^{229}{\rm Th}$ to isomeric states through electron recollision \cite{Wang:2021,QiPRL2023}, and can influence the half-life of $\alpha$ decay by modifying the Coulomb barrier \cite{Delion:2017,Qi:2019,Ghinescu:2020,Palffy:2020,Qi:2020}, etc.

The exploration of fusion cross section for light nuclei under intense laser fields has attracted extensive attention owing to its potential application in clean energy sources \cite{Queisser:2019,Lv:2019,Wang:2020,Liu:2021,Kohlfu:2021,Lv:2022,
Qi:2022,Wu20221,Bekx20221,Bekx20222,liu2022}.
Previous studies \cite{Queisser:2019,Lv:2019,Wang:2020,Liu:2021,Kohlfu:2021,Lv:2022,
Qi:2022,Bekx20221,Bekx20222,liu2022} mainly utilize the celebrated
Gamow form \cite{Gamow:1928,Atzeni:2004} for fusion cross section and focus on the enhancement effects on tunneling probability through the Coulomb repulsive potential. However, these approaches lack a self-consistent description of nuclear potential combined with Coulomb repulsion potential in intense laser fields. Given the intricate nature  of the nuclear potential, the phenomenological optical potentials within mean-field approximation find extensive use in describing nuclear potential \cite{Dickhoff:2018,lixingzhong:2000,lixingzhong:2008,Singh:2019,Wu20222}.
A complex spherical square-well optical potential is commonly employed to characterize nuclear potentials in the context of light nuclear fusion \cite{lixingzhong:2000,lixingzhong:2008,Singh:2019,Wu20222},
wherein the imaginary component of the potential signifies the decay of the compound nucleus. In our recent work \cite{Wu20221}, applying a  complex spherical square-well model and exploiting Kramers's approximation, we observed
the distinct shift in the peaks of fusion cross-section within strong high-frequency laser fields, attributed to resonant tunneling mechanism. Intense low-frequency laser fields, particular in the near-infrared regime, are generated by a majority of advanced laser facilities worldwide \cite{Zhang2020}. In scenarios of intense low-frequency fields, a substantial number of photons (e.g., $>10000$) are involved. These laser photons are extensively employed to investigate the multiphoton ionization of atomic and molecular systems \cite{Joachain2012,Liu:2014}.
The question of whether multiphoton processes can affect the light nuclear cross section remain unresolved.

In this study, we extend  our recent theoretical work \cite{Wu20221} to  the low-frequency regime, where a complex spherical square well \cite{lixingzhong:2000,lixingzhong:2008,Singh:2019,Wu20222} is employed to characterize the nuclear potential. Utilizing the Coulomb-Volkov solutions which are successfully employed in handing nonperturbative multiphoton process in the electron-atom collision \cite{Ehlotzky1998}, we self-consistently derive the multiphoton cross sections of nuclear fusion. Our specific focus is on deuteron-triton (DT)  and proton-boron (p~$^{11}{\rm B}$) fusion. DT fusion takes precedence in controlled fusion research owing to its relatively high fusion cross section compared to other fusion reactions \cite{Atzeni:2004}. On the other hand, p~$^{11}{\rm B}$ fusion reaction is an aneutronic process involving abundant and stable isotopes \cite{Lv20222,Tentori2023,Istokskaia2023}. The potential applications of this reaction span from controlled nuclear fusion to an emerging form of proton-boron capture therapy \cite{Tentori2023,Istokskaia2023}. Our findings reveal that for longer laser wavelengths and lower incident particle kinetic energies, fusion cross sections exhibit substantial enhancement for both DT and p~$^{11}{\rm B}$ fusion reactions due to multiphoton processes.

The structure of this work is outlined as follows:
Section II introduces our theoretical framework for multiphoton fusion. Sec. III presents our main results and discussions. Sec. IV provides the main conclusions and outlooks.

\section{Theoretical framework}

In this section, we present a theoretical model for examining multiphoton fusion in the presence of a background laser field. The laser field is assumed to be a monochromatic wave with frequency $\omega$  and linear polarized along the $z$ axis. Under the dipole approximation,
it is described by the uniform and time varying vector potential $\bm{A}(t)=A_0{\rm cos}(\omega t)\vec{e}_z$,
where $A_0$ is the amplitude.

\begin{figure}[!tb]
\centering
\includegraphics[width=\linewidth]{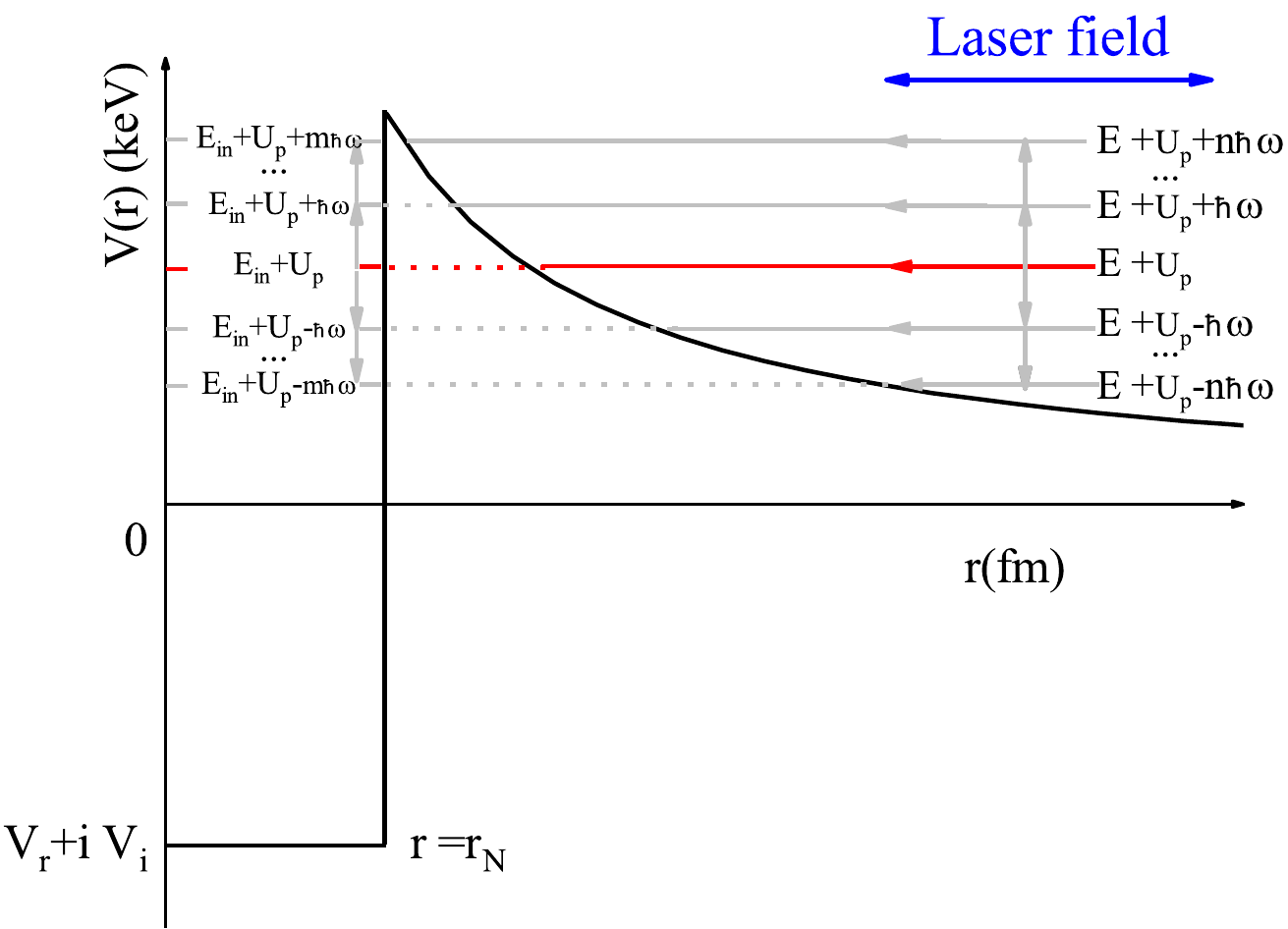}
\caption{(Color online) A schematic representation of the model for multiphoton fusion processes in
the presence of a linearly polarized laser field. The potential $V(r)$ is considered to be a short-range complex
spherical square well with a long-range Coulomb repulsive potential between two nuclei.
 The charge particle within a intense laser field exhibits a range of possible energies due to multiphoton processes.
Here, $E$ represents the kinetic energy of fusion
system in the CM frame, $U_p$ denotes the ponderomotive energy of the particle in the laser fields, and $n$ corresponds the number of photons absorbed or emitted outside nuclear potential. $E_{\rm in}$ is the energy within nuclear potential, and $m$ represents the number of photons absorbed or emitted inside nuclear potential. }
\label{figure1}
\end{figure}

In the presence of laser fields, the relative motion of a two-body, spinless fusion system in the center-of-mass (CM) frame can be described by the time-dependent Schr\"{o}dinger equation
\begin{equation}\label{TDSE}
i\hbar\frac{\partial}{\partial t}\Psi(t,{\bm r})=\left[\frac{1}{2\mu}\left(\hat{{\bm P}}-q_{\rm eff} {\bm A}(t)\right)^2+V(r)\right]\Psi(t,{\bm r}),
\end{equation}
where $\mu=m_{\rm p} m_{\rm t}/(m_{\rm p}+m_{\rm t})$ is reduced mass of two nuclei. $m_{\rm p}$ and $m_{\rm t}$ are masses of incident projectile and target nuclei, respectively. $q_{\rm eff}=e(Z_{\rm p} m_{\rm t} -Z_{\rm t} m_{\rm p})/(m_{\rm p}+m_{\rm t})$ is an effective charge, where $Z_{\rm p}$ and $Z_{\rm t}$ are charge numbers  of incident projectile and target nuclei, respectively. As illustrated in Fig. \ref{figure1}, the potential $V(r)$ of the fusion process is characterized by a short-range complex square nuclear potential with a long-range Coulomb repulsive potential between two nuclei. The expression for $V(r)$ is
given by
\begin{equation}\label{potential}
V(r)=\left\{
\begin{array}{rcl}
V_{\rm r}+ i V_{\rm i},& & {r < r_{\rm N},}\\
\frac{ e^2}{4\pi \epsilon_0 r},& & {r > r_{\rm N},}\\
\end{array} \right.
\end{equation}
where $r_{\rm N}$ is the radius of nuclear well. $V_{\rm r}$ and $V_{\rm i}$
can describe the scattering and absorption effects during the  fusion, respectively, which be known as the ``optical model".
Comparing with experimental benchmark cross section data in the absence of electromagnetic fields, the three optical parameters $V_{\rm r}$, $V_{\rm i}$ and $r_{\rm N}$ for light fusion reactions have been calibrated in our recent work \cite{Wu20222}.

\subsection{Solution by expansion on spherical Coulomb-Volkov states}

We attempt to solve the time-dependent Schr\"{o}dinger equation (\ref{TDSE}) by the expansion on spherical Coulomb-Volkov states. According to the quantum scattering theory, the wave function $\Psi(t,{\bm r})$ after being
scattered  by the ansatz
can be written as
\begin{eqnarray}\label{Psi11}
\Psi(t,{\bm r})&=&\Psi_{\rm inc}(t,{\bm r})+\Psi_{\rm scatt}(t,{\bm r})\nonumber\\
&=&\sum_{l=0}^{+\infty}(2l+1)i^l\frac{{\rm F}_l(k_nr,\eta_n)}{k_nr}P_l(\vartheta)e^{-\frac{i}{\hbar}Et}e^{-\frac{i}{\hbar}
\int^t_{-\infty}V_L{\rm d}\tau}\nonumber\\
&+&\sum_{n=-\infty}^{+\infty}\sum_{l=0}^{+\infty}(2l+1)i^lA_{n,l}\frac{{\rm H}_l(k_n r,\eta_n)}{k_n r}\nonumber\\
&\times&P_l(\vartheta)e^{-\frac{i}{\hbar}E_n t},r>r_N,
\end{eqnarray}
where ${\rm F}_l$ is regular Coulomb wave function \cite{Abramowitz1972}, $P_l$ is the Legendre polynomials, $\vartheta$ is the polar angle in the spherical coordinate, $E$ is the kinetic energy of fusion system in the CM frame, $V_L$ is the interaction potential between the laser field and the nucleus, respectively. In the intense laser field, as shown in Fig. \ref{figure1}, the charge particle has a series of possible energies $E_n=E+U_p+n\hbar\omega$ for all integers $n$, where $n$ is the number of photons absorbed or emitted and $U_p=q^2_{\rm eff}A^2_0/(4\mu)$ is the ponderomotive energy of the particle in the laser fields, respectively. Accordingly, the particle wave number $k_n=\sqrt{2\mu E_n/\hbar^2}$ and dimensionless Coulomb parameter $\eta_n=1/(k_na_c)$ in the laser fields, respectively,
where $a_{\rm c}=4\pi\epsilon_0\hbar^2/(\mu Z_{\rm p} Z_{\rm t} e^2)$ is Coulomb unit length. ${H_l}=F_l-iG_l$, where $G_l$ is irregular Coulomb wave function \cite{Abramowitz1972}.

In view of temporal integrals in the exponent inside Eq. (\ref{Psi11}), by inserting vector potential ${\bm A}(t)$ into Eq. (\ref{Psi11}) and exploiting the Jocabi-Anger identity \cite{Abramowitz1972}, one can obtain
\begin{eqnarray}\label{exp}
e^{-\frac{i}{\hbar}\int^t_{-\infty}V_L{\rm d}\tau}&=&e^{-\frac{i}{\hbar}\int^t_{-\infty}
\left(-\frac{q_{\rm eff}}{\mu}{\bm A}(t)\cdot{\bm p}+\frac{q^2_{\rm eff}{\bm A}^2}{2\mu}\right){\rm d}\tau}\nonumber\\
&=&e^{i\left(\zeta{\rm sin}\omega t-\frac{z}{2}{\rm sin}2\omega t-U_p t\right)}\nonumber\\
&=&\sum^{+\infty}_{-\infty}\tilde{J}_n\left(\zeta,-\frac{z}{2}\right)
e^{-\frac{i}{\hbar}(U_p+n\hbar\omega)t},
\end{eqnarray}
where $\zeta=q_{\rm eff}A_0p{\rm cos}\theta/(\hbar\omega)$, $z=q_{\rm eff}A_0/(4\mu\hbar\omega)$, $\tilde{J}_n$ is the generalized Bessel function \cite{Abramowitz1972}. $\theta$ is the angle between the particle momentum ${\bm P}$ and laser polarization axis (the $z$ axis).
Putting Eq. (\ref{exp}) into Eq. (\ref{Psi11}), $\Psi(t,{\bm r})$ can be further written as
\begin{eqnarray}\label{Psi12}
\Psi(t,{\bm r})
&=&\sum_{n=-\infty}^{+\infty}\sum_{l=0}^{+\infty}(2l+1)i^l
\tilde{J}_n\left(\zeta,-\frac{z}{2}\right)\frac{{\rm F}_l(k_nr,\eta_n)}{k_nr}\nonumber\\
&\times&P_l(\vartheta)e^{-\frac{i}{\hbar}E_n t}\nonumber\\
&+&\sum_{n=-\infty}^{+\infty}\sum_{l=0}^{+\infty}(2l+1)i^lA_{n,l}\frac{{\rm H}_l(k_n r,\eta_n)}{k_n r}\nonumber\\
&\times&P_l(\vartheta)e^{-\frac{i}{\hbar}E_nt},r>r_N.
\end{eqnarray}

Similarly, the wave function $\Psi(t,{\bm r})$ that satisfies the boundary conditions inside nuclear potential well can be given by
\begin{eqnarray}
\Psi(t,{\bm r})
&=&\sum_{m,E_{\rm in}}\sum_{l=0}^{+\infty}B_l(E_{\rm in})(2l+1)i^l
\tilde{J}_n\left(-\zeta,\frac{z}{2}\right)\frac{{\rm j}_l(k_N r,\eta)}{k_Nr}\nonumber\\
&\times&P_l(\vartheta)e^{-\frac{i}{\hbar}E_m t}, r<r_N£¬
\end{eqnarray}
where $B_l(E_{\rm in})$ is a coefficient, $E_{\rm in}$ is the energy in nuclear potential,
 $j_l$ is the Riccati-Bessel function \cite{Abramowitz1972}, $k_N=\sqrt{2\mu(E_{\rm in}-V_{\rm r}-iV_{\rm i})/\hbar^2}$ is the complex wave number in the nuclear well.

\subsection{The continuity conditions of wave function at the radius of nuclear well}

The coefficients $A_{n,l}$ inside Eq. (\ref{Psi12}) are so far unknown and would be determined by the means of the the continuity conditions of wave function at the radius of nuclear well.
 The continuity conditions of wave function and its first derivative are satisfied simultaneously at $r=r_N$ for
all values of $\vartheta$ and $t$, one can obtain
\begin{eqnarray}\label{continuity}
E+U_p+m\hbar\omega&=&E_{\rm in}+U_p+n\hbar\omega\nonumber\\
\sum_m B_l(E_{\rm in})j_m(k_Nr_N)\tilde{J}_n&=&{\rm F}_l(k_n r_N)\tilde{J}_n+A_{n,l}{\rm H}_l(k_n r_N)\nonumber\\
\sum_mB_l(E_{\rm in})k_Nj^\prime_m(k_Nr_N)\tilde{J}_n&=&k_n\left[{\rm F}^\prime_l(k_n r_N)\tilde{J}_n+A_{n,l}{\rm H}^\prime_l(k_nr_N)\right].\nonumber\\
\end{eqnarray}
One can easily find that $E_{\rm in}=E+(n-m)\hbar\omega$ and $k_{N,m}=\sqrt{2m(E+(n-m)\omega-V_{\rm r}-iV_{\rm i})/\hbar^2}$, respectively.
We can define the logarithmic derivative
\begin{eqnarray}
C_l\equiv\frac{\sum_m B_l(E+(n-m)\omega)k_{N,m}j^\prime_l(k_Nr_N)\tilde{J}_m}{\sum_m B_l(E+(n-m)\omega)j_l(k_Nr_N)\tilde{J}_m}.
\end{eqnarray}
We assume $B_l(E+(n-m)\hbar\omega)\approx B_l(E+n\hbar\omega)$ and
according to Eq. (\ref{continuity}), one can then find
\begin{eqnarray}
A_{n,l}&=&\frac{\left[C_{n,l}{\rm F}_l(k_nr_N,\eta)-k_n{\rm F}^\prime_l(k_n r_N,\eta_n)\right]\tilde{J}_n\left(\zeta,-\frac{z}{2}\right)}{k_n{\rm H}^\prime_l(k_n r_N,\eta_n)-C_{n,l}{\rm H}_l(k_n r_N,\eta_n)}\nonumber\\
&\equiv&A^0_{n,l}\tilde{J}_n\left(\zeta,-\frac{z}{2}\right).
\end{eqnarray}

\subsection{Multiphoton fusion cross sections }

For the evaluation of fusion cross sections in a laser field, we start from the asymptotic form of the wave function $\Psi(t,{\bm r})$ [i.e., Eq. (\ref{Psi12})]. Since for $r\rightarrow+\infty$, ${\rm F}_l(k_nr,\eta_n)\rightarrow[e^{i(k_nr-l\pi/2)}-e^{-i(k_nr-l\pi/2)}]/(2i)$, ${\rm H}_l(k_n r,\eta_n)\rightarrow-ie^{i(k_n r-l\pi/2)}$, the wave function $\Psi({\bm r})$ may be written in asymptotic form as
\begin{eqnarray}\label{asymptotic}
\Psi({\bm r}\rightarrow+\infty)
&=&\sum_{n=-\infty}^{+\infty}\sum_{l=0}^{+\infty}(2l+1)(-1)^{l+1}\left[
\tilde{J}_n\frac{e^{-ikr}}{2ikr}\right.\nonumber\\
&+&\left.(2A_{n,l}+\tilde{J}_n)\frac{e^{ik_nr}}{2ik_n r}
\right]\times P_l(\vartheta).
\end{eqnarray}
The probability current of the particles in a laser field reads
\begin{eqnarray}
{\bm J}_{\rm c}(t)=\frac{\hbar}{2im}(\Psi^\ast\nabla\Psi-\Psi\nabla\Psi^\ast)
-\frac{q}{\mu}{\bm A}(t)\Psi^\ast\Psi.
\end{eqnarray}
The time-average current during a period of laser field can be written as
\begin{eqnarray}
\bar{{\bm J}}_{\rm c}=
\frac{1}{T}\int_0^T {\bm J}_{\rm c}(t){\rm d}t
=\frac{\hbar}{2i\mu}(\Psi^\ast\nabla\Psi-\Psi\nabla\Psi^\ast).
\end{eqnarray}
The time-average current of incoming particles, denoted as $\bar{{\bm J}}_{\rm c}^{{\rm inc}}$, is computed from the incoming wave function as
\begin{eqnarray}
\bar{{\bm J}}_{\rm c}^{{\rm inc}}=\frac{\hbar}{2i\mu}(\Psi_{\rm inc}^\ast\nabla\Psi_{\rm inc}-\Psi_{\rm inc}\nabla\Psi_{\rm inc}^\ast)=\frac{\hbar {\bm k}_n}{\mu}.
\end{eqnarray}
The radial component of the time-average current in the laser field is essential for fusion cross section and it reads
\begin{eqnarray}\label{current}
\bar{J}_{\rm c}^r=\bar{{\bm J}}_{\rm c}\cdot \hat{e}_r
=\frac{\hbar}{2i\mu}(\Psi^\ast\frac{\partial}{\partial r} \Psi-\Psi\frac{\partial}{\partial r}\Psi^\ast).
\end{eqnarray}
Putting Eq. (\ref{asymptotic}) into Eq. (\ref{current}), we can find
\begin{eqnarray}
\bar{J}_{\rm c}^r=\sum_{n=-\infty}^{+\infty}\bar{J}_{\rm c}^{rn}&=&\sum_{n=-\infty}^{+\infty}\sum_{l,l_1}\frac{(2l+1)(2l_1+1)}{4k_n\mu r^2}P_l(\vartheta)P_{l_1}(\vartheta)\nonumber\\
&\times&\left\{(2A^\ast_{n,l}+\tilde{J}_n)(2A_{n,l}+\tilde{J}_n)-(-1)^{l+l_1}
\tilde{J}_n^2\right\}.\nonumber\\
\end{eqnarray}

The differential fusion cross sections in each solid angle can be defined
as
\begin{eqnarray}
\frac{{\rm d}\sigma(E,\theta,\vartheta)}{{\rm d}\Omega}=\sum_{n=-\infty}^{+\infty}\frac{{\rm d}\sigma_n(E,\theta,\vartheta)}{{\rm d}\Omega}=\sum_{n=-\infty}^{+\infty}
-\frac{\bar{J}_{\rm c}^{rn}r^2{\rm d}\Omega}{\bar{J}_{\rm c}^{{\rm inc}}{\rm d}\Omega}.\nonumber\\
\end{eqnarray}
By the integrating over the solid angle ${\rm d}\Omega$, we can easily obtain the total fusion cross sections:
\begin{eqnarray}\label{crosssection}
\sigma(E,\theta)&=&\sum_{n=-\infty}^{+\infty}\sigma_n(E,\theta)=\sum_{n=-\infty}^{+\infty}
\int\frac{{\rm d}\sigma_n(E,\vartheta,\theta)}{{\rm d}\Omega} {\rm sin\vartheta {\rm d}\vartheta}\nonumber\\
&=&\sum_{n,l=-\infty}^{+\infty}
\frac{(2l+1)\pi\left(\tilde{J}_n^2-|2A_{n,l}+\tilde{J}_n|^2
\right)}{k_n^2}\nonumber\\
&=& \sum_{n,l=-\infty}^{+\infty}\frac{(2l+1)\pi\left(1-|2A^0_{n,l}+1|^2
\right)}{k_n^2}\tilde{J}_n^2\left(\zeta,-\frac{z}{2}\right)\nonumber\\
&\equiv& \sum_{n,l=-\infty}^{+\infty}\sigma_l(E_n)P_n(\theta).
\end{eqnarray}
Equation (\ref{crosssection}) has a clear physical interpretation, representing the product of the cross section $\sigma_l(E_n)$ associated with the possible energy $E_n$ of a charged particle and the conresponding probability $P_n(\theta)=\tilde{J}_n^2\left(\zeta,-\frac{z}{2}\right)$ of the particle absorbing or emitting photons in intense laser fields. It is noteworthy that $\sum_n P_n(\theta)=1$.

\section{Numerical results and discussions}

In the context of the low-energy ($E<1~{\rm MeV}$)~fusion reaction in the following calculations, our focus is only on computing the $S$ wave ($l=0$) contribution to the cross section. In a low-frequency intense laser field, the energy of laser photon (on the order of eV) is significantly less than the both absolute value of the real part $|V_{\rm r}|$ (on the order of MeV)
of optical potential and the collision energy $E$~(on the order of keV) so that the complex wave number in the nuclear well
$k_{N,m}\approx \sqrt{2\mu(E-V_{\rm r}-iV_{\rm i})/\hbar^2}$. This result implies that low-frequency intense field primarily alters the energy of fusion particles located outside the nuclear well.
\begin{figure}[!t]
\centering
\includegraphics[width=\linewidth]{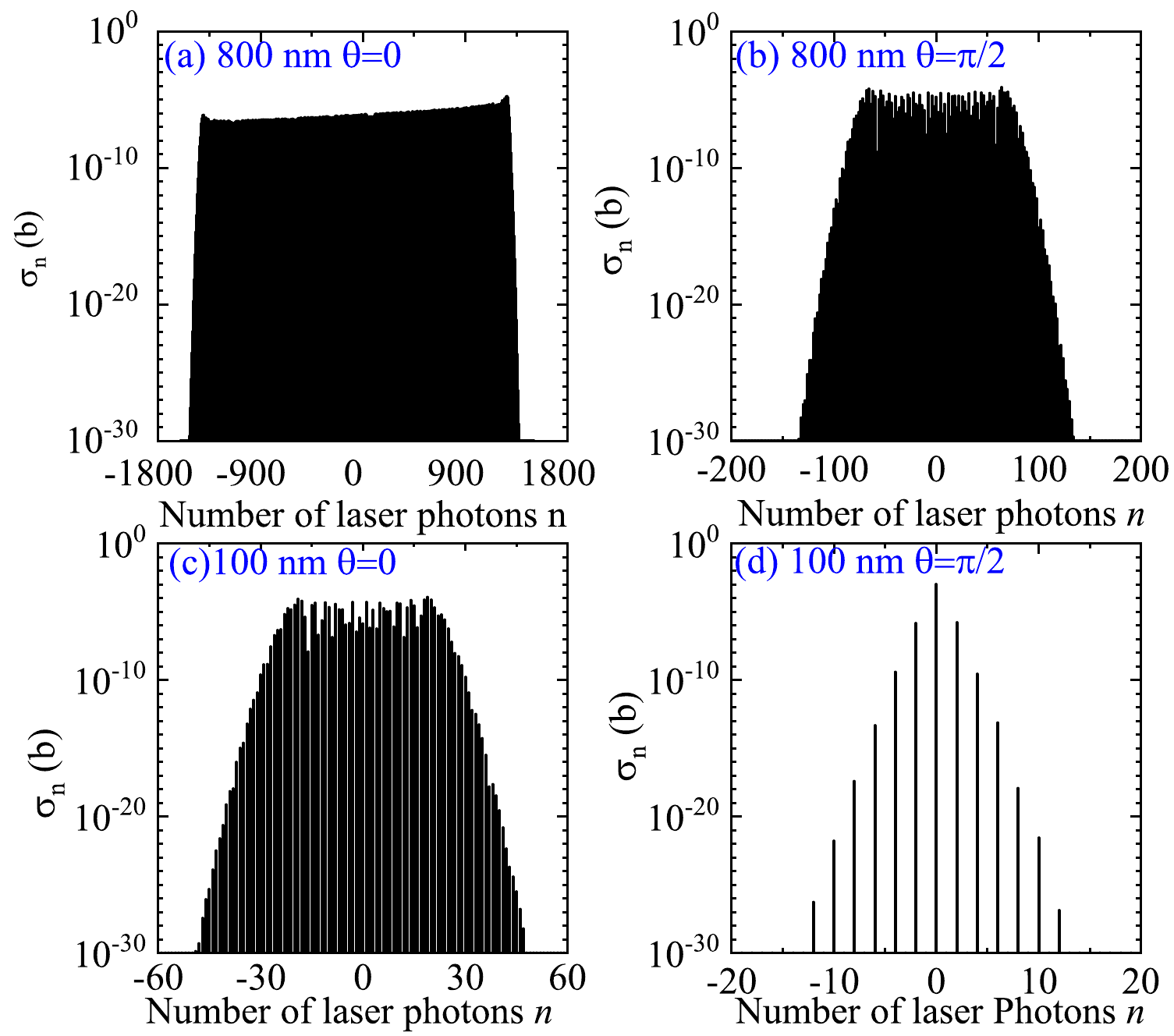}
\caption{(Color online)(a)-(d) Distributions of the multiphoton cross sections $\sigma_n$ for laser-assisted DT fusion as a function of the number of absorbed ($n>0$) or emitted ($n<0$) laser photons. The laser intensity is $10^{20}{\rm W/cm^2}$ and the relative kinetic energy of DT nuclei $E=5~{\rm keV}$ for all the two laser wavelengths 800nm (top row) and 100nm (bottom row). For each wavelength, two
$\theta$ are shown as labeled in figure.}
\label{figure2}
\end{figure}

\subsection{Multiphoton fusion cross sections of DT fusion}

\subsubsection{Distributions of multiphoton fusion cross sections}

We first examine the distributions of multiphoton fusion cross sections $\sigma_n$ concerning the number of absorbed ($n>0$) or emitted ($n<0$) laser photons for DT fusion.  Examples of $\sigma_n$ are presented in Fig. \ref{figure2} for wavelengths of 800 nm and 100 nm, employing a consistent
laser intensity of $10^{20}{\rm W/cm^2}$ and the relative kinetic energy of $E=5~{\rm keV}$. Two angles $\theta$ are
indicated for each wavelength. The number of laser photons involved in fusion reaction is highly dependent on both the laser wavelength and the angle between the incident direction of the particle and the laser polarization axis.
As seen in Figs. \ref{figure2} (a) and (c), as the laser wavelength shortens, the number of photons involved  decrease from more over 2000 photons (800 nm) to the only 120 photons (100 nm). This is partly due to the fact that shorter wavelength result in fewer photons per unit area for the fixed laser intensity. On the other hand,
Figs. \ref{figure2} (a)-(b) demonstrate that as the angle changes from inclination angle $\theta=0$ (i.e.,the incident direction of the particle is parallel to the laser polarization direction) to $\theta=\pi/2$ (i.e., the direction perpendicular to the polarization direction), the number of laser photons involved becomes smaller, while the peak values of the multiphoton cross sections increase.
Notably, understanding the presence of laser photons
at $\theta=\pi/2$ proves challenging within the classical framework.

Moreover, Figs. \ref{figure2}~(a)-(c) reveal a distinct nonperturbative plateau structure: the multiphoton cross sections decrease rapidly when the number of absorbed or emitted laser photons exceeds a certain threshold. In conclusion, we highlight that total cross sections $\sigma=\sum_n\sigma_n$ in Figs. \ref{figure2}~(a)-(d) are approximately $6.5\times10^{-4}$ b, $2.7\times10^{-4}$ b, $2.3\times10^{-4}$ b and $2.3\times10^{-4}$ b, respectively.
Surprisingly, the results for the laser wavelength 100nm (Figs. \ref{figure2}~(c)-(d)) indicate that the presence of the laser field contributes the
the photo-particle energies over a broad range, but it hardly affects the total cross sections compared with the field-free
case~($2.3\times10^{-4}$ b). It is worth noting that a similar phenomenon also occurs in the distributions
of the multiphoton cross sections for laser-assisted proton emission in Ref. \cite{Dadi2012}.

\subsubsection{Angle-averaged effective fusion cross section }
\begin{figure}[!t]
\centering
\includegraphics[width=0.8\linewidth]{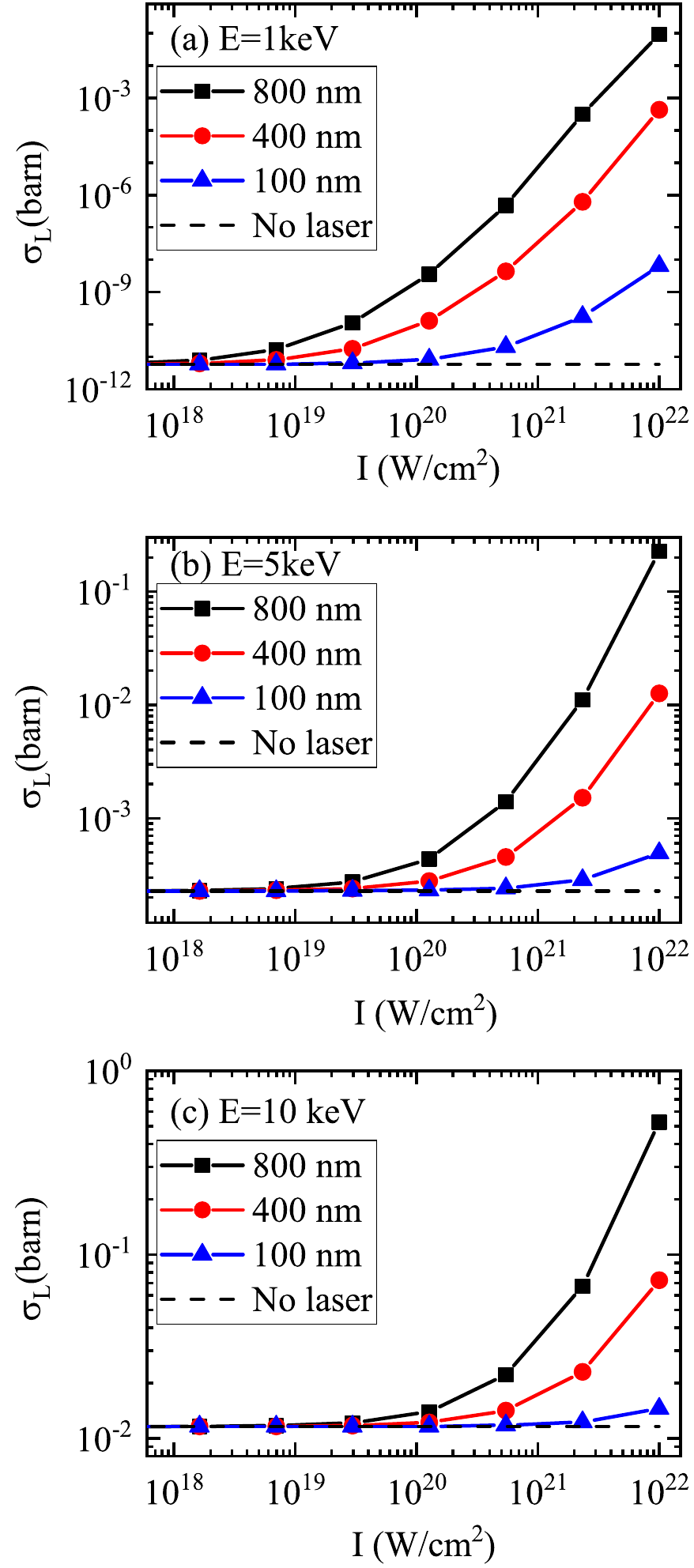}
\caption{(Color online) Angle-averaged effective DT fusion cross-section $\sigma_L$ under different laser intensities and wavelengths for $E$=(a) 1 keV, (b) 5 keV, and (c) 10 keV. The horizontal dashed line in each figure marks the corresponding laser-free cross section.}
\label{figure3}
\end{figure}
In the thermal environment of nuclear fusion, the angle between the incident particle wave number ${\bm k}$ and laser polarization axis is random. To provide a more quantitative measure , the angle-averaged effective fusion cross section in the laser fields can be expressed as \cite{Ghinescu:2020,Wang:2020}
\begin{eqnarray}
\sigma_L(E)=\frac{1}{2}\int_0^{\pi}\sigma(E,\theta){\rm sin}\theta {\rm d}\theta.
\end{eqnarray}
We display  angle-averaged effective cross section $\sigma_L$ of DT fusion in Fig. \ref{figure3} under different laser intensities and wavelengths. These laser
parameters are anticipated in intense laser facilities worldwide \cite{Yoon,http:eli,Zhang2020}. Three specific kinetic energies $E$=(a) 1 keV, (b) 5 keV and (c) 10 keV in Fig. \ref{figure3} are relevant to controlled fusion research \cite{Atzeni:2004}. The horizontal dashed line in each figure marks the corresponding laser-free cross section.

One can observe in Fig. \ref{figure3} that, for all the three energies, the $\sigma_L$ are substantially enhanced compared to the corresponding laser-free case,  with intensities on the order of $10^{22}{\rm W/cm^2}$. Moreover,
Fig. \ref{figure3} illustrates that with longer laser wavelength and the low relative kinetic energy $E$, the laser field can more efficiently transfer energy to the DT fusion system, resulting in higher $\sigma_L$. For instance, at a laser wavelength of 800 nm and a relative kinetic energy of 1 keV, the $\sigma_L$ increases by approximately ten orders of magnitude. However, for a laser wavelength of 100 nm and  a the relative kinetic energy of 10 keV, the $\sigma_L$ only exhibits a one-fold increase.


\begin{figure}[!t]
\centering
\includegraphics[width=\linewidth]{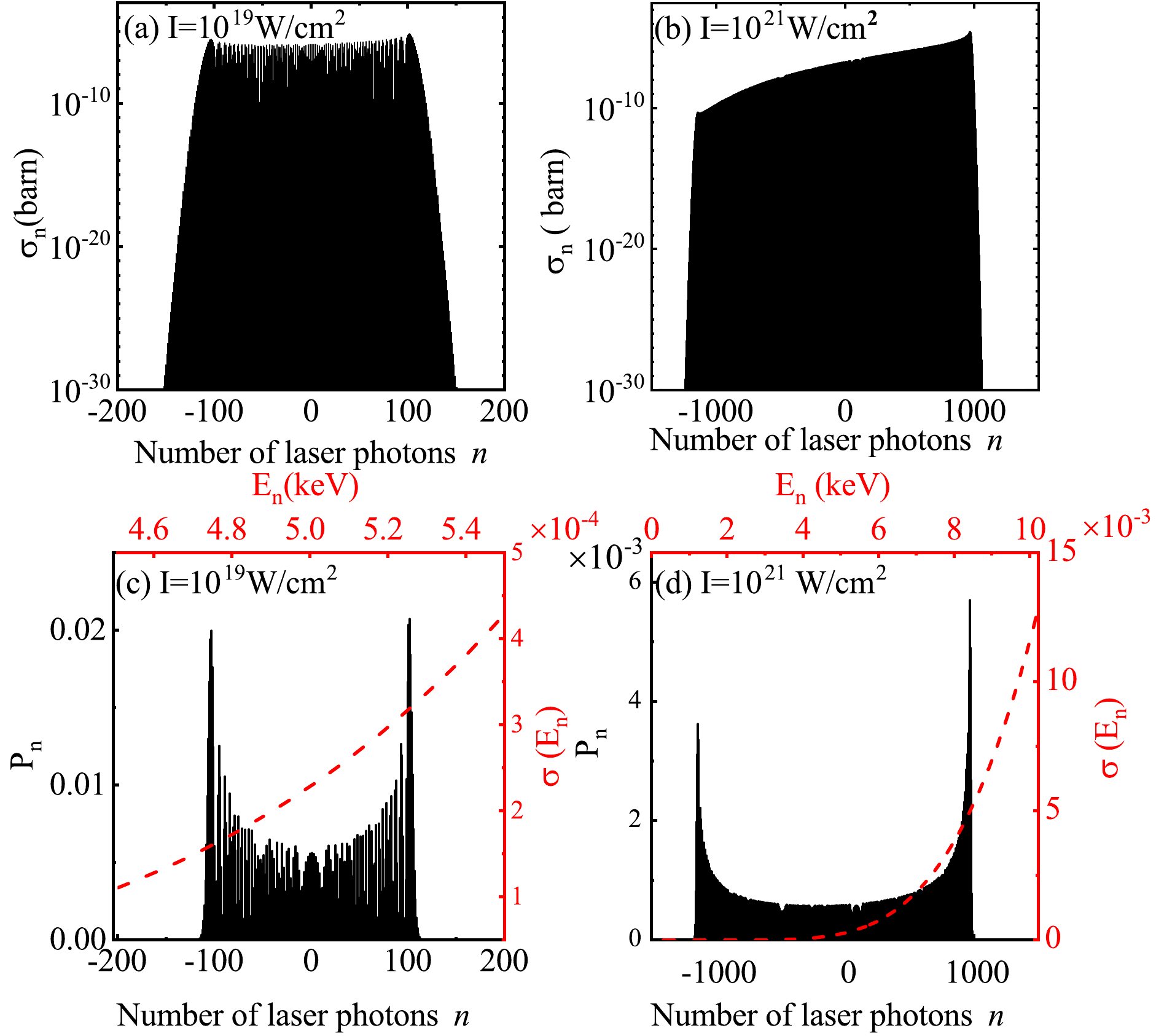}
\caption{(Color online) Distributions of the multiphoton cross sections $\sigma_n$ as a function of the number of absorbed ($n>0$) or emitted ($n<0$) laser photons for two intensities (a) $I=10^{19}{\rm W/cm^2}$ and (b) $10^{21}{\rm W/cm^2}$,respectively. (c)-(d): Corresponding distributions of  the probability $P_n$ (black solid lines) and cross section $\sigma(E_n)$
(red dot lines). The laser wavelength, the relative kinetic energy of DT nuclei
and the incident angle are 400nm, $5~{\rm keV}$ and 0, respectively.}
\label{figure4}
\end{figure}

To gain a deeper understanding of the enhanced fusion cross sections with increasing laser intensity, we focus on two
scenarios: (a) $I=10^{19}{\rm W/cm^2}$ and (b) $I=10^{21}{\rm W/cm^2}$ in  Fig. \ref{figure4} for detailed discussions. Other fixed parameters, including the
the laser wavelength (400 nm), the relative kinetic energy of DT nuclei (5 keV), and the incident angle (0), remain constant. We present the distributions of multiphoton cross sections as a function of the number of laser photons in Figs. \ref{figure4} (a)-(b). It is evident from these figures that as the lase intensity increases, the number of laser photons involved in DT fusion process significantly  increase. Specifically, at a laser intensity of $I=10^{19}{\rm W/cm^2}$, three hundred laser photons involved in fusion process, while at $I=10^{21}{\rm W/cm^2}$, more than two thousand laser photons are involved. Additionally, the multiphoton distribution exhibits a pronounced asymmetry with increasing laser intensity.

Simultaneously, we present the corresponding distributions of the probability $P_n$ (black solid lines) and cross section $\sigma(E_n)$ (red dot lines) in Figs. \ref{figure4} (c)-(d) for possible energy $E_n$ of the particle under intense laser field. It is crucial to note that multiphoton cross section $\sigma_n$ is calculated as the product of probability $P_n$ and cross section $\sigma(E_n)$
associated with the possible energy $E_n$, as defined by Eq. (\ref{crosssection}).

In Fig. \ref{figure4} (c), it is evident that as the number of absorbed or emitted photons increases, the probability $P_n$ also increases. The symmetrical structure of $P_n$ with respect to $n=0$ in Fig. \ref{figure4} (c) indicates that the probability of absorbed photons in fusion process is equivalent to the probability of emitting photons. Additionally, Fig. \ref{figure4} (c) reveals that at a laser intensity of $I=10^{19}{\rm W/cm^2}$, the possible energy of fusion particle involved in fusion process is
$4.7~{\rm keV}<E_n<5.3~{\rm keV}$. Correspondingly,
the cross section $\sigma(E_n)$ exhibits a monotonic and gradual increase $E_n$, resulting in $\sum_n\sigma_n(E,\theta)\approx\sigma(E)\sum_n P_n(\theta)=\sigma(E)$.
Consequently, in Fig. \ref{figure3} (c), when the laser wavelength and intensity are 400nm and $10^{19}{\rm W/cm^2}$, the enhancement of fusion cross section in laser field is almost negligible.

However, with an increase in laser intensity ($ I=10^{21}{\rm W/cm^2}$), the probability distribution $P_n$ in Fig. \ref{figure4} (d) displays a pronounced left-right asymmetry concerning $n=0$ : the probability of absorbing photons exceeds that of emitted photons for a large number of photons. Additionally, compared to the scenario in Fig. \ref{figure4} (c), the possible energy $E_n$ of the particles involved in fusion process have a large range
from $1~{\rm keV}$ to $9~{\rm  keV}$. The corresponding cross section $\sigma(E_n)$ also exhibits a rapid and
monotonic increase with $E_n$. The increased number of involved photons and higher probability of absorbing laser photons contribute to an enhanced fusion cross section under laser fields with a wavelength of 400 nm and an intensity of
$I=10^{21}{\rm W/cm^2}$ in Fig. \ref{figure3} (b). In summary, the rise in laser intensity leads to a monotonic
increase in both the number of laser photons involved in fusion process and the corresponding absorption probability,
resulting in a monotonically enhanced cross section in Fig. \ref{figure3}.

\subsection{Multiphoton cross sections of p~$^{11}{\rm B}$ fusion}
\begin{figure}[!t]
\centering
\includegraphics[width=0.8\linewidth]{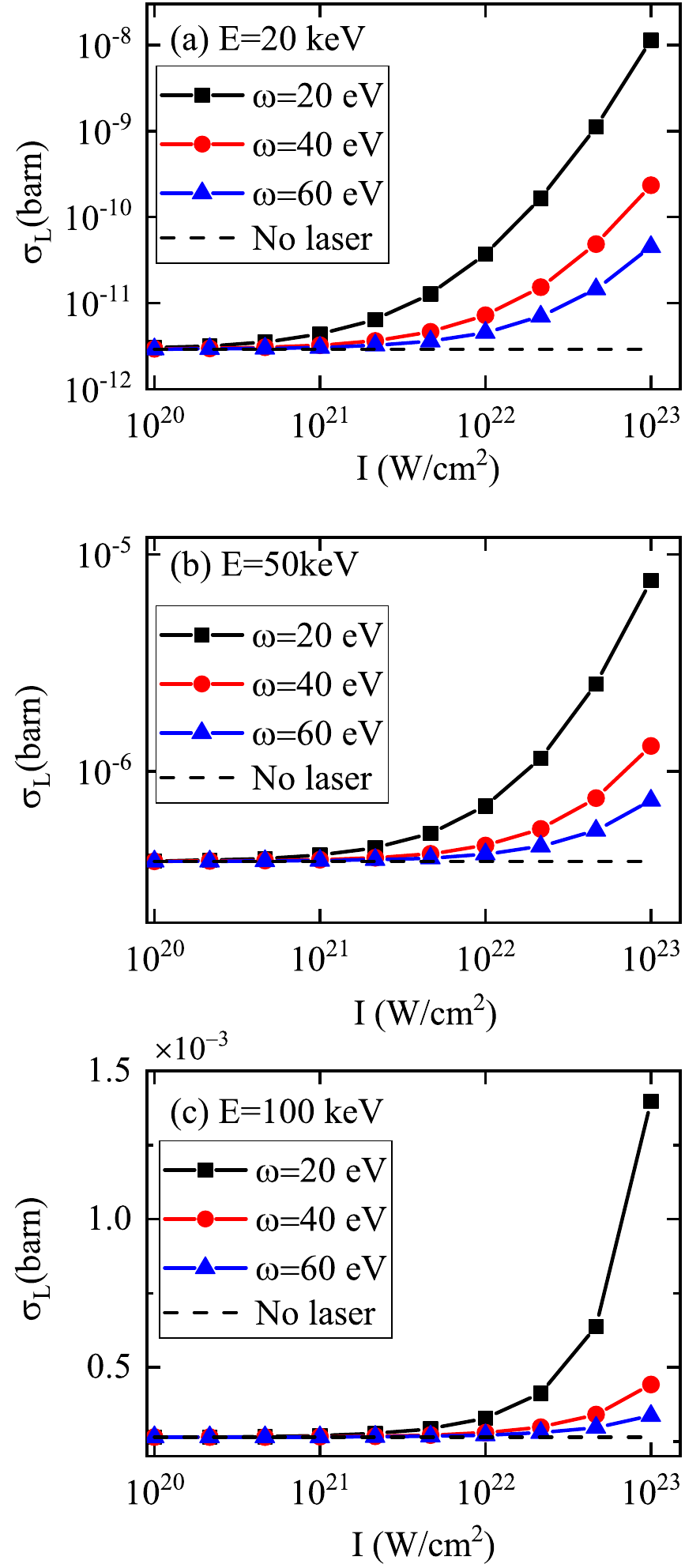}
\caption{(Color online) Angle-averaged effective P~$^{11}{\rm B}$ fusion cross-section $\sigma_L$ under different laser intensities and wavelengths for $E$=(a) 20 keV, (b) 50 keV, and (c) 100 keV. The horizontal dashed line in each figure marks the corresponding laser-free cross section.}
\label{figure5}
\end{figure}
In this section, our focus is on the p~$^{11}{\rm B}$ fusion reaction. We have calibrated three optical parameters of
 $V_{\rm r}$, $V_{\rm i}$ and $r_{\rm n}$ for p~$^{11}{\rm B}$ fusion at low energy using precise Coulomb wave functions \cite{Wu20222}. Subsequently,
 we investigate the multiphoton effect of p~$^{11}{\rm B}$ fusion. Distributions of multiphoton cross sections for p~$^{11}{\rm B}$ fusion resemble those of DT
 fusion and they are not presented here. Angle-averaged effective  cross sections $\sigma_L$ of p~$^{11}{\rm B}$ fusion  are displayed in Fig. \ref{figure5} under various laser intensities and wavelengths for the relative kinetic energy $E$=(a) 20 keV, (b) 50 keV, and (c) 100 keV, respectively. The horizontal dashed line in each figure indicates the corresponding laser-free cross section.

The data presented in Fig. \ref{figure5} indicate that, for all the three kinetic energies,  $\sigma_L$ are enhanced compared to the corresponding laser-free case under high laser intensities. Similar to DT fusion, when considering a longer laser wavelength and low the kinetic energy $E$, the $\sigma_L$ for  p~$^{11}{\rm B}$ fusion exhibits substantial increases in comparison to the laser-free case. Specifically, at a constant laser intensity I=$10^{23}{\rm W/cm^2}$, the $\sigma_L$ exhibits a nearly four-order-of-magnitude increase for a laser wavelength of 800 nm and kinetic energy of 20 keV. In contrast, for a laser wavelength of 100 nm and kinetic energy of 100 keV, the increase is only a 0.2-fold, as depicted in Fig. \ref{figure5}. Furthermore, compared to DT fusion results,
higher laser intensities are necessary to influence the cross section of p~$^{11}{\rm B}$ fusion reaction due to the elevated Coulomb repulsion barriers.

\section{Conclusions and outlooks}

In conclusion, we have examined the physics of multiphoton
fusion reactions involving light nuclei such as DT and p~$^{11}{\rm B}$. This investigation incorporates a combination of the Coulomb-Volkov solutions and the complex spherical square-well nuclear potential. Our theoretical
findings reveal that, particularly with longer laser wavelengths and lower incident particle kinetic energies,
several thousand photons can participate in the fusion reactions, resulting in a significant enhancement of fusion cross section. Additionally, we observe distinct nonperturbative plateau structures in the distributions of multiphoton cross sections. These outcomes suggest that intense low-frequency lasers could prove instrumental in advancing controlled fusion research, potentially relaxing the temperature requirements outlined in the well-known Lawson criterion \cite{Lawson1957}

In this study, we exclusively examine the low-frequency regime of laser fields. The $\gamma$ ray-assisted multiphoton fusion processes involving nuclear excitation merit further exploration. Furthermore, it is essential to acknowledge that the complex spherical square well represents the most basic optical model for describing the
nuclear potential during fusion. Future investigations
should contemplate the utilization of a more realistic
optical potential featuring a rigid core and nuclear spin,
such as the Woods-Saxon forms \cite{Newton2004}, particularly in context of multiphoton fusion processes.

\section*{Acknowledgments}

This work was supported by funding from NSAF No. U2330401, and NSFC No. 12375235 and No. 12174034.

\end{document}